\documentclass[useAMS,usenatbib]{mn2e}
\usepackage{graphicx}
\usepackage{amssymb}

\title[The spectral lag of short and long GRBs]{Comparing the spectral lag of short and long gamma-ray bursts and its relation with the luminosity}

\author[Bernardini et al.]{M. G. Bernardini$^{1}$\thanks{E--mail:grazia.bernardini@brera.inaf.it}, G. Ghirlanda$^1$, S. Campana$^{1}$, S. Covino$^{1}$, R. Salvaterra$^2$, \newauthor
J.-L. Atteia$^{3,4}$, D. Burlon$^5$, G. Calderone$^1$, P. D'Avanzo$^{1}$, V. D'Elia$^{6,7}$, \newauthor
G. Ghisellini$^{1}$, V. Heussaff$^{3,4}$, D. Lazzati$^8$, A. Melandri$^{1}$, L. Nava$^{9}$, \newauthor 
S. D. Vergani$^{10}$ and G. Tagliaferri$^{1}$\\
$^{1}$ INAF-Osservatorio Astronomico di Brera, via E. Bianchi 46, I-23807 Merate, Italy\\
$^{2}$ INAF-IASF Milano, via E. Bassini 15, I-20133 Milano, Italy \\
$^{3}$ Universit\'e de Toulouse; UPS-OMP; IRAP; Toulouse, France\\
$^{4}$ CNRS IRAP, 14 avenue Edouard Belin, F-31400 Toulouse, France\\
$^{5}$ Sydney Institute for Astronomy, School of Physics, The University of Sydney, NSW 2006, Australia \\
$^{6}$ INAF-Osservatorio Astronomico di Roma, via Frascati 33, I-00040 Monteporzio Catone (RM), Italy\\
$^{7}$ ASI-Science Data Center, Via del Politecnico snc, I-00133 Rome, Italy \\
$^{8}$ Department of Physics, Oregon State University, 301 Weniger Hall, Corvallis, OR 97331, USA\\
$^{9}$ Racah Institute of Physics, The Hebrew University of Jerusalem, 91904, Israel\\
$^{10}$ GEPI, Observatoire de Paris, CNRS, Univ. Paris Diderot, 5 place Jules Janssen, 92190, Meudon, France\\
}
\date{\today}

\pagerange{\pageref{firstpage}--\pageref{lastpage}} \pubyear{2013}

\begin{document}

\label{firstpage}

\maketitle

\begin{abstract}
We investigated the rest frame spectral lags of two complete samples of bright long ($50$) and short ($6$) gamma-ray bursts (GRB) detected by \textit{Swift}. We analysed the \textit{Swift}/BAT data through a discrete cross-correlation function (CCF) fitted with an asymmetric Gaussian function to estimate the lag and the associated uncertainty. We find that half of the long GRBs have a positive lag and half a lag consistent with zero. All short GRBs have lags consistent with zero. The distributions of the spectral lags for short and long GRBs have different average values. Limited by the small number of short GRBs, we cannot exclude at more than $2\,\sigma$ significance level that the two distributions of lags are drawn from the same parent population. If we consider the entire sample of long GRBs, we do not find evidence for a lag-luminosity correlation, rather the lag-luminosity plane appears filled on the left hand side, thus suggesting that the lag-luminosity correlation could be a boundary. Short GRBs are consistent with the long ones in the lag-luminosity plane. 
\end{abstract}


\section{Introduction}\label{sect_i}

Extensive studies on BATSE gamma-ray bursts (GRB) found evidences that global spectral evolution within the prompt emission is a general trend for \textit{long}\footnote{\textit{Long} GRBs are conventionally defined as those with $T_{90}\geqslant 2$ s, while \textit{short} GRBs are those with $T_{90} < 2$ s. This definition was first derived for BATSE GRBs in the $25-300$ keV energy band \citep{1993ApJ...413L.101K}. $T_{90}$ is the time interval during which the central $90\%$ of all counts are recorded by the detector.} GRBs \citep{1986ApJ...301..213N,1986AdSpR...6...19N,1995A&A...300..746C,1996ApJ...459..393N,1995ApJ...439..307F,1997ApJ...486..928B,2002A&A...393..409G}. In particular, \citet{1995A&A...300..746C} identified a delay in the arrival times of low-energy photons with respect to high-energy photons. This \textit{spectral lag} is conventionally defined positive when high-energy photons precede low-energy photons. \citet{2000ApJ...534..248N} showed that the positive spectral lag anti-correlates with the burst bolometric peak luminosity within a limited sample of BATSE long GRBs with redshift measurements. Since then, this anti-correlation has been further investigated and confirmed with different samples of BATSE \citep{2002ApJ...579..386N,2007ApJ...660...16S,2008ApJ...677L..81H}, HETE \citep{2010PASJ...62..487A} and \textit{Swift} bursts \citep{2010ApJ...711.1073U}. An anti-correlation between the spectral lag and the bolometric peak luminosity similar to the one derived for the prompt emission has been found for X-ray flares \citep{giantflares10}.

\citet{2012MNRAS.419..614U} recently performed the first detailed analysis of the spectral lag for long GRBs observed by \textit{Swift} \citep{2005ApJ...621..558G} with the Burst Alert Telescope (BAT, \citealt{2005SSRv..120..143B}). With respect to previous studies, the availability of many redshifts allowed them to compute the lags in two selected \textit{rest frame energy bands} for all the considered burst. They confirmed the existence of the correlation with a smaller scatter when compared to previous analyses in the observer-frame \citep[e.g.][]{2010ApJ...711.1073U}. However, in the determination of the lag-luminosity correlation they did not consider $44\%$ of the GRBs of their original sample that have spectral lag consistent with zero or negative.

The existence of the spectral lag has been interpreted either as a consequence of the spectral evolution \citep{1998ApJ...501L.157D,2003ApJ...594..385K,2005A&A...429..869R,2011AN....332...92P}, or as due to the curvature effect, i.e. related to the delay in the arrival time of photons emitted at high latitude with respect to the observer line of sight \citep{2000ApJ...544L.115S,2001ApJ...554L.163I,2004ApJ...614..284D,2005MNRAS.362...59S,2006MNRAS.367..275L}, or it may reveal the existence of separate spectral components evolving independently \citep{2010ApJ...725..225G,2013ApJ...770...32G}. Practically, the spectral lag has been used as a possible tool to discriminate between long and short GRBs \citep{2006Natur.444.1044G}, since the latter tend to have a smaller lag (consistent with zero) with respect to long GRBs \citep{2001grba.conf...40N,2006ApJ...643..266N}.

In this paper we provide a comprehensive analysis of the spectral lag for both long and short GRBs. We make use of the complete sub-samples of bright GRBs observed by BAT presented in \citet[][long GRBs]{2012ApJ...749...68S} and in \citet[][short GRBs]{2014arXiv1405.5131D} to constrain the properties of the spectral lag. Since the spectral lag value is dependent upon the energy bands chosen to compute it for both short \citep{2009Natur.462..331A,2010ApJ...725..225G,2013ApJ...770...32G} and long GRBs \citep{2010ApJ...711.1073U}, we adopted two fixed rest-frame energy bands to perform a direct comparison of the lags of the two classes of long and short GRBs. Our method is similar to that adopted by \citet{2012MNRAS.419..614U}, but we account for the possible asymmetry of the cross-correlation function when computing the lag and its uncertainty. We also verified our findings applying our method to a \textit{Fermi} Gamma Burst Monitor (GBM) sample. We investigate the lag-luminosity relation accounting also for the large fraction of long GRBs with lag consistent with zero. 

In Section~\ref{sect_m} we discuss the method we adopted to extract the spectral lag. In Section~\ref{sect_ss} we describe the sample selection criteria. In Section~\ref{sect_r} we show our results about the spectral lags derived for short and long GRBs and the lag-luminosity relation. In Section~\ref{sect_c} we draw our main conclusions. We adopted standard values of the cosmological parameters: $H_\circ=70$ km s$^{-1}$ Mpc$^{-1}$, $\Omega_M=0.27$, and $\Omega_{\Lambda}=0.73$. Errors are given at the $1\, \sigma$ confidence level unless otherwise stated.

\section{Methodology}\label{sect_m}

\begin{figure}
\centering
\includegraphics[width= \hsize,clip]{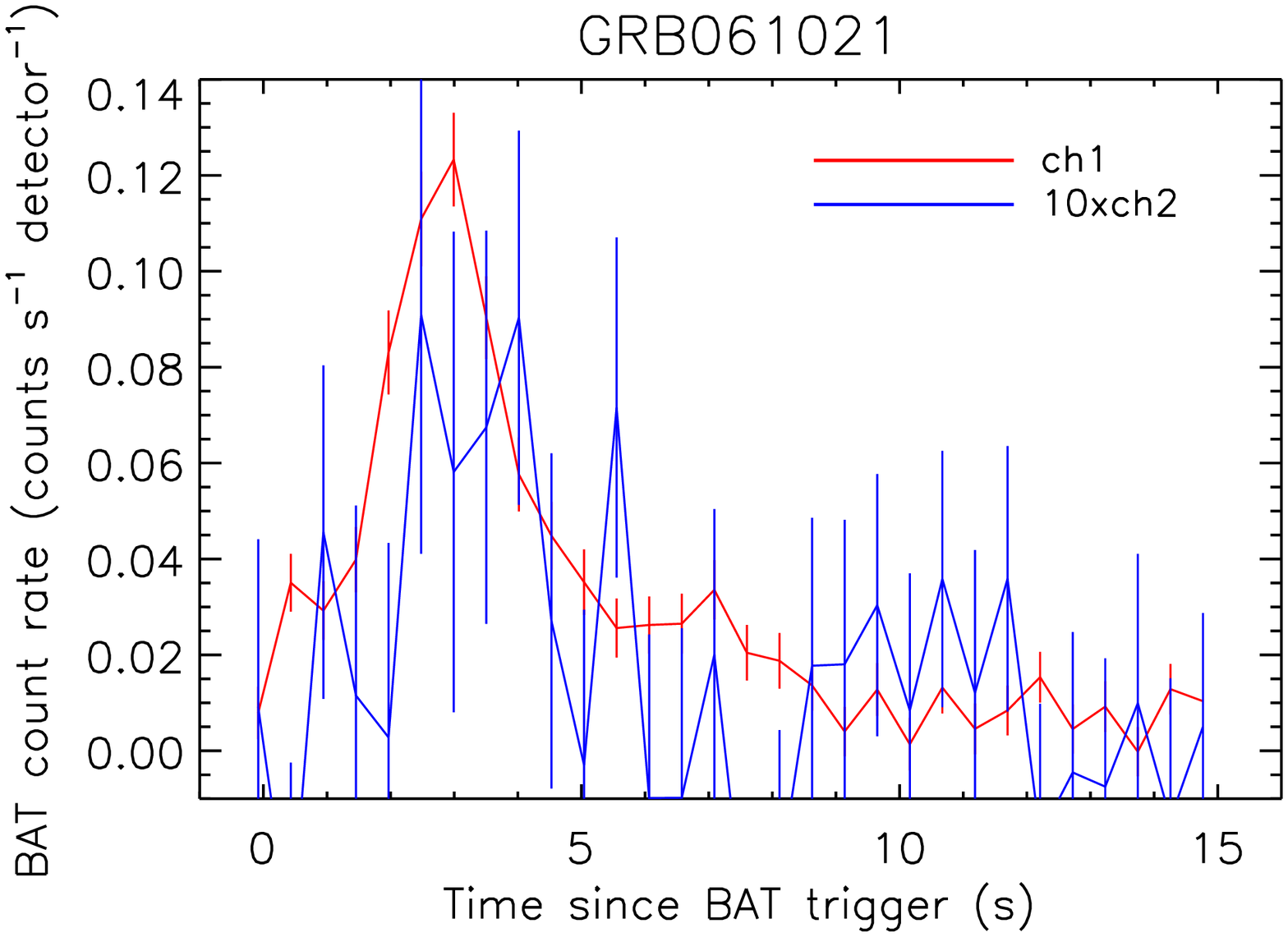}\\
\includegraphics[width= \hsize,clip]{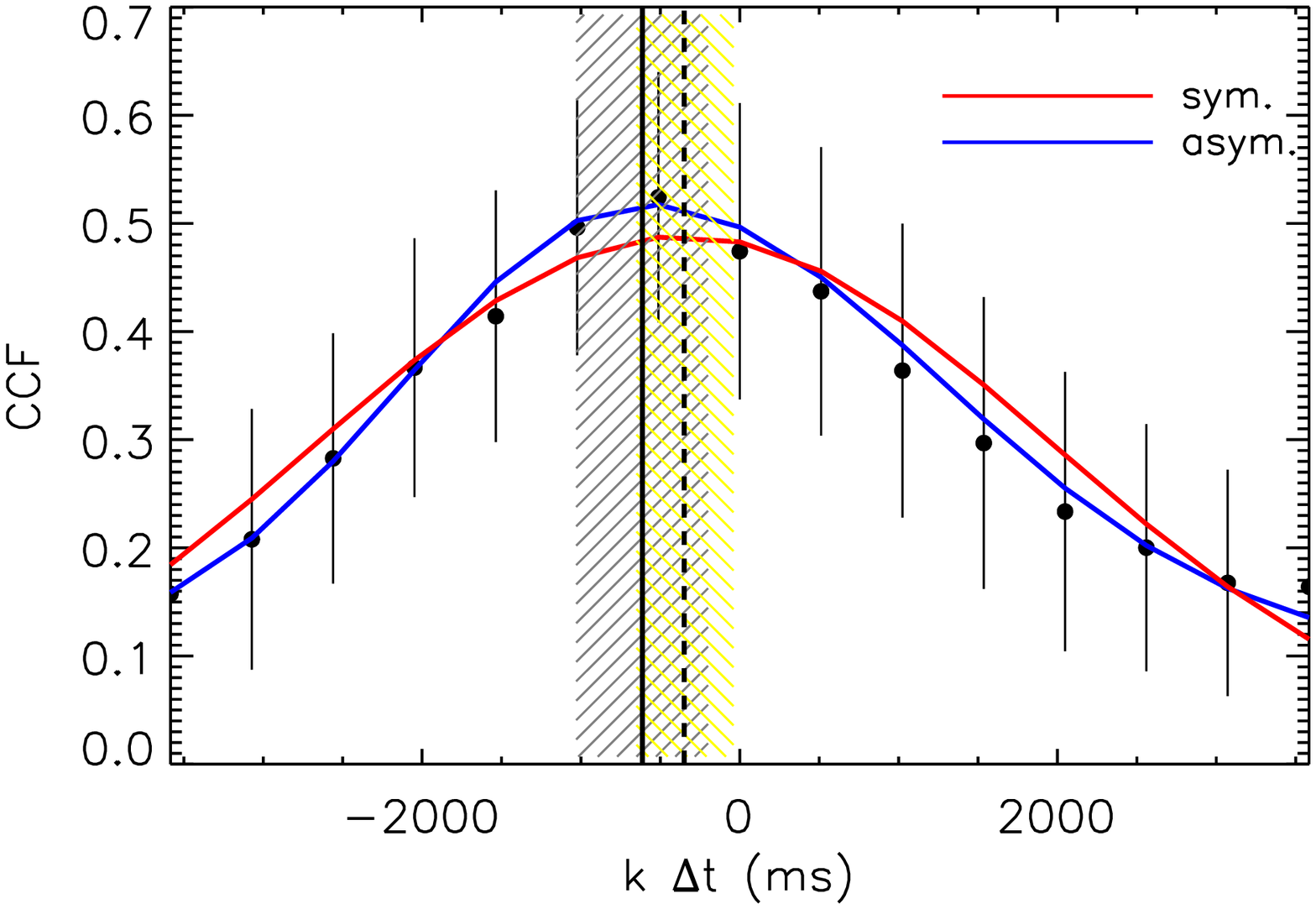}
\caption{Lag computation for GRB 061021. Upper panel: BAT count rate light curve in the $100-150$ keV rest frame energy band (ch1, red points) and in the $200-250$ keV rest frame energy band (ch2 multiplied by a factor $10$, blue points), binned with $\Delta t=512$ ms in the time interval selected for the lag computation ($[0-15]$ s). Lower panel: Cross-correlation function (CCF) calculated for the light curves in ch1 and ch2 (black points). The errors are obtained through a flux-randomization method. The blue line represents the best fit to the CCF with an asymmetric Gaussian model while the red line shows the best fit with a symmetric Gaussian model. The solid vertical line corresponds to the maximum for the asymmetric Gaussian model, and the dashed vertical line corresponds to the maximum for the symmetric Gaussian model. The gray and yellow hatched areas mark the $1\,\sigma$ uncertainties for the asymmetric and symmetric Gaussian models, respectively.}
\label{fig_061021}
\end{figure}

We retrieved \textit{Swift}/BAT data from the public archive\footnote{http://heasarc.gsfc.nasa.gov/cgi-bin/W3Browse/swift.pl} and processed them with the standard \textit{Swift} analysis software included in the NASA's HEASARC software (HEASOFT, ver. 6.15.1) and the relevant latest calibration files. For each GRB, we extracted mask-weigthed, background-subtracted light curves with the \texttt{batmaskwtevt} and \texttt{batbinevt} tasks in FTOOLS for two observer-frame energy bands corresponding to the fixed rest-frame energy bands $100-150$ keV (hereafter ch1) and $200-250$ keV (hereafter ch2). Raw light curves (non mask-weighted) can be contaminated by other sources and by background variations with time, due to the slewing of the spacecraft during the prompt emission and are severely affected by extra variance, which is comparable with the Poisson variance due to the counting statistics, thus, are not suitable for temporal variability studies \citep{2007MNRAS.379..619R}. The choice of the same rest-frame energy bands that have been adopted in the recent work by \citet{2012MNRAS.419..614U} allows us to make a direct comparison of our results. Besides, the GRBs in our samples (long and short) cover a redshift range $z\in[0.35-5.47]$, thus the observer-frame energy range covered in our analysis (i.e. $[100-150]/(1+z_{max})$ and $[200-250]/(1+z_{min})$) are within the energy range of the BAT instrument ($\sim [15-200]$ keV; \citealt{2011ApJS..195....2S}).

In order to measure the temporal correlation of the two light curves in ch1 and ch2 we used the discrete cross-correlation function (CCF). Differently from the standard Pearson CCF, we adopted a modified CCF non-mean subtracted derived by  \citet{1997ApJ...486..928B}, that is more suitable for transient events such as GRBs:
\begin{equation}
{\rm CCF}(k\Delta t;c_1,c_2)=\frac{\sum_{i=max(1,1-k)}^{min(N,N-k)}\, {c_1}_i {c_2}_{(i+k)}}{\sqrt{\sum_i\,{{c_1}_i}^2\, \sum_i\, {{c_2}_i}^2}}\, ,
\end{equation}
where $\Delta t$ is the duration of the time bin, whose choice is described below, $k\Delta t$ ($k=..,-1,0,1,...$) is a multiple of the time bin and represents the time delay,  $c_1$ and $c_2$ are the count rates of ch1 and ch2, respectively. Here the summation is extended over the total number of data points $N$ considered in the light curve. For each GRB we select the time interval (see Table~\ref{table_tot}) over which both the light curves of ch1 and ch2 have the main emission episode, avoiding to include, for example, long-lasting tails or long quiescent times, and we calculated the CCF over that time interval to avoid that the CCF technique associates unrelated structures.

For each light curve pairs, we calculated the CCF value for a series of time delays $k\Delta t$ and we defined the spectral lag $\tau$ as the time delay that corresponds to the global maximum of the CCF versus time delay: CCF$(\tau) = $ max[CCF$(k\Delta t; c_1, c_2)$]. 

To locate the global maximum, we fit an asymmetric Gaussian model to the CCF versus time delay:
\begin{eqnarray}
{\rm CCF}(x)= {\rm const} + \Bigg\{\begin{array}{c} N {\rm exp}\left[-\frac{(x-\tau)^2}{2\Sigma_l^2} \right]\, \,x\leqslant \tau \\ N {\rm exp}\left[-\frac{(x-\tau)^2}{2\Sigma_r^2} \right]\, \, x> \tau \end{array} \, .
\end{eqnarray}
The use of a continuous function fit to the discrete CCF and search for its maximum allows us to estimate lags $\tau$ which can be a fraction of the time resolution $\Delta t$ of the light curves extracted from the BAT data. 

We choose an asymmetric Gaussian model since it reflects the natural asymmetry of the CCF inherited by the asymmetry of the GRB pulses \citep{1997ApJ...486..928B}. In order to support our choice, we simulated single-peaked synthetic light curves introducing an artificial lag and we fitted to the CCF both a symmetric and an asymmetric Gaussian model. Though both fits are statistically acceptable, the asymmetric model systematically recovers the real value, while the symmetric one tends to overestimate it. Besides, the asymmetric model is not alternative to the symmetric case but includes it under the condition $\Sigma_l=\Sigma_r$, therefore using the asymmetric Gaussian model allows us to be more general than in the symmetric case and to better represent the shape of the CCF in those cases where it can be derived with high precision. In fig.~\ref{fig_061021} we show the lag computation for GRB 061021 as an example, comparing the fit to the CCF (black points in the bottom panel of fig.~\ref{fig_061021}) performed with an asymmetric and a symmetric Gaussian function.

The uncertainties on the CCF have been derived by applying a flux-randomization method \citep{1998PASP..110..660P}: we generated $10,000$ realizations of ch1 and ch2 light curves based on each count rate $\bar{c_i}$ and its error $\Delta c_i$: $c_i=\bar{c_i}+\xi_i \times \Delta c_i$, where $\xi_i$ is a random number drawn from a standard normal distribution\footnote{By virtue of the mask weighting technique, BAT mask-weighted light curves and spectra have gaussian statistics.}. For each time delay $k\Delta t$, the corresponding value of the CCF and its error are the mean and the standard deviation of the distribution of the CCF calculated from the $10,000$ realizations of the light curves. We derived the uncertainty on the spectral lag from the fit to $1,000$ different realizations of the CCF versus time delay with the same randomization method. These Montecarlo simulations allows us to provide a reasonable estimate of the uncertainties compared to the temporal bin $\Delta t$, as also proved by \citet{2010ApJ...711.1073U}.

We successfully tested this procedure with synthetic light curves where we introduced artificial lags. Both the accuracy and the precision of the lag extraction depend on the signal-to-noise ratio (S/N) of the light curves \citep{1997ApJ...486..928B,2010ApJ...711.1073U}. For noisy light curves, the CCF maximum value decreases, and the CCF versus time delay is much more scattered, leading to a less accurate determination of the spectral lag (see \citealt{2010ApJ...711.1073U}, therein Fig.~2). For this reason, when working with real data, we started with mask-weighted light curves binned at $\Delta t=4$ ms and we progressively doubled the bin size until we found that the chance probability of finding such a value of CCF$_{\rm max}$ with the number of data points considered is $< 10^{-3}$. In this way we also avoid to consider statistical fluctuations as the global maximum. At the same time, the use of a continuous function fit to the CCF allows us to estimate lags $\tau$ which can be a fraction of the time binning of the light curves, thus no strong bias is introduced in the lag estimates by using different temporal bins. We report in Table~\ref{table_tot} for each GRB the time resolution adopted in the light curves of ch1 and ch2 to compute the spectral lag.

\section{Sample selection}\label{sect_ss}

We selected two samples of long and short GRBs observed by \textit{Swift}/BAT and with a secure redshift determination. The redshift is needed  to properly select the common rest frame energy range for the lag $\tau$ computation (as described in Sect.~\ref{sect_m}) and to correct the spectral lag itself for the cosmological time dilation effect, i.e. $\tau_{RF}=\tau/(1+z)$. Moreover, since we want to explore the reliability of the spectral lag-luminosity correlation, we need the redshift to compute the isotropic equivalent luminosity $L_{\rm iso}=4\pi d_{\rm L}(z)^{2} F$, from the rest frame bolometric ($1-10^4$ keV) peak flux $F$ (here $d_{\rm L}(z)$ is the luminosity distance corresponding to the redshift $z$). To the latter aim we need that the prompt emission spectrum has been observed over a wide energy range and that its peak energy (of the  $\nu F_{\nu}$ spectrum) has been measured. However, this requirement is not introduced as a selection criterion of the samples: we first select the sample of short and long GRBs with known redshifts and compute their rest frame lags $\tau_{RF}$ and then we explore the lag-luminosity correlation for those bursts with also a secure measurement of their $L_{\rm iso}$. 

For the long GRBs we made use of the complete sub-sample of bright long GRBs observed by BAT presented in \citet{2012ApJ...749...68S} (hereafter BAT6), comprising $54$ GRBs with secure redshift determination (for the latest compilation of the sample see \citealt{2013arXiv1303.4743C}). The choice of a complete flux-limited sample of bursts ensures to study the lag properties with a sample which is less affected by instrumental selection effects, though being limited in number because composed only by the bright end of the population of {\it Swift} GRBs. However, as extensively discussed in several papers \citep{2012ApJ...749...68S,2012MNRAS.421.1697C,2012MNRAS.425..506D,2012MNRAS.421.1265M,2012MNRAS.421.1256N,2013arXiv1303.4743C,2013MNRAS.435.2543G}, the BAT6 sample is suited to study the prompt and afterglow properties of GRBs being almost free from instrumental selection effects: this sample contains all the bursts with favorable observing conditions from ground that \textit{Swift} has detected above a flux limit of 2.6 ph cm$^{-2}$ s$^{-1}$ (in the $15-150$ keV energy band) and, a posteriori, it has a high degree of completeness in redshift ($95\%$).

We applied the procedure described in Sect.~\ref{sect_m} to the long GRBs of the BAT6 sample and we ended up with $50$ long GRBs with a significative global maximum in the CCF ($4$ GRBs from the BAT6 sample - GRB 050416A, GRB 060614, GRB 071112C and GRB 081007 - have been discarded from our analysis because there were not enough counts in one of the two energy bands - usually ch2 - to extract a significant value of the CCF for any choice of the time bin). Specifically, we determined:
\begin{itemize}
\item $25$ GRBs with positive lag (i.e. not consistent with zero within $1\,\sigma$; $50\%$ of the sample),
\item $2$ with negative lag ($4\%$)
\item $23$ with lag consistent with zero ($46\%$), being its central value either positive ($10$; $20\%$) or negative ($13$; $26\%$). 
\end{itemize}
These results are reported in Table~\ref{table_tot}. 

A similar analysis has been recently presented by \citet{2012MNRAS.419..614U} with an incomplete sample of bright long GRBs observed by BAT and with a slightly different methodology. Indeed, the major difference from \citet{2012MNRAS.419..614U} is that they use a symmetric Gaussian model fit to the data, while we accounted for the possible asymmetry of the CCF as a result of the intrinsic asymmetry of the pulse profiles composing GRB light curves (see fig.~\ref{fig_061021}). We checked if this modification introduces systematic effects in the lag determination by comparing the rest-frame lags extracted for $30$ long GRBs common to the sample adopted in \citet{2012MNRAS.419..614U}. Overall, we find consistency between the two methods in all cases, as expected since the symmetric case is included in the asymmetric one, with a tendency of having smaller uncertainties for the asymmetric model in the cases with high statistics.

\begin{figure}
\centering
\includegraphics[width= \hsize,clip]{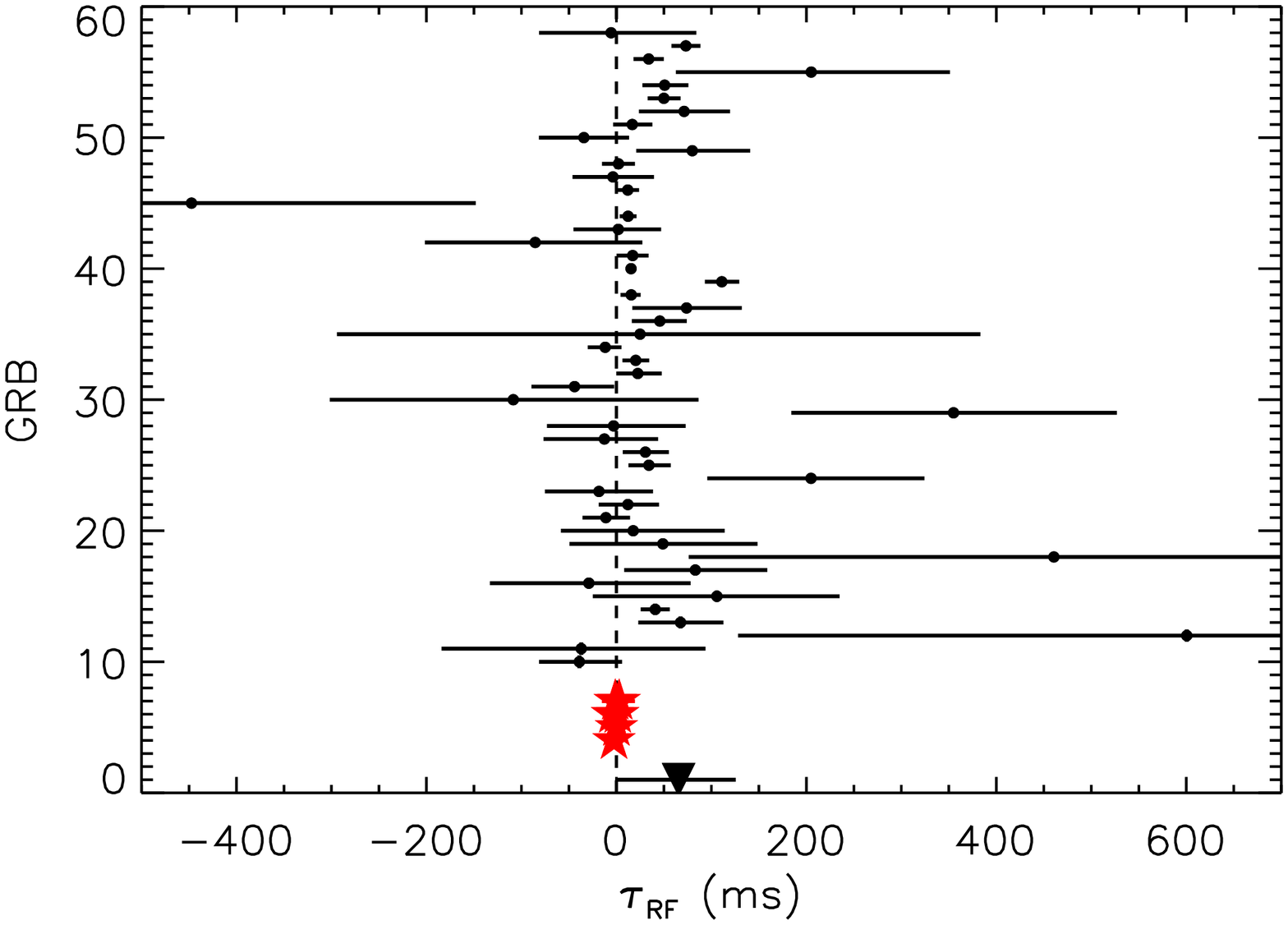}\\
\includegraphics[width= \hsize,clip]{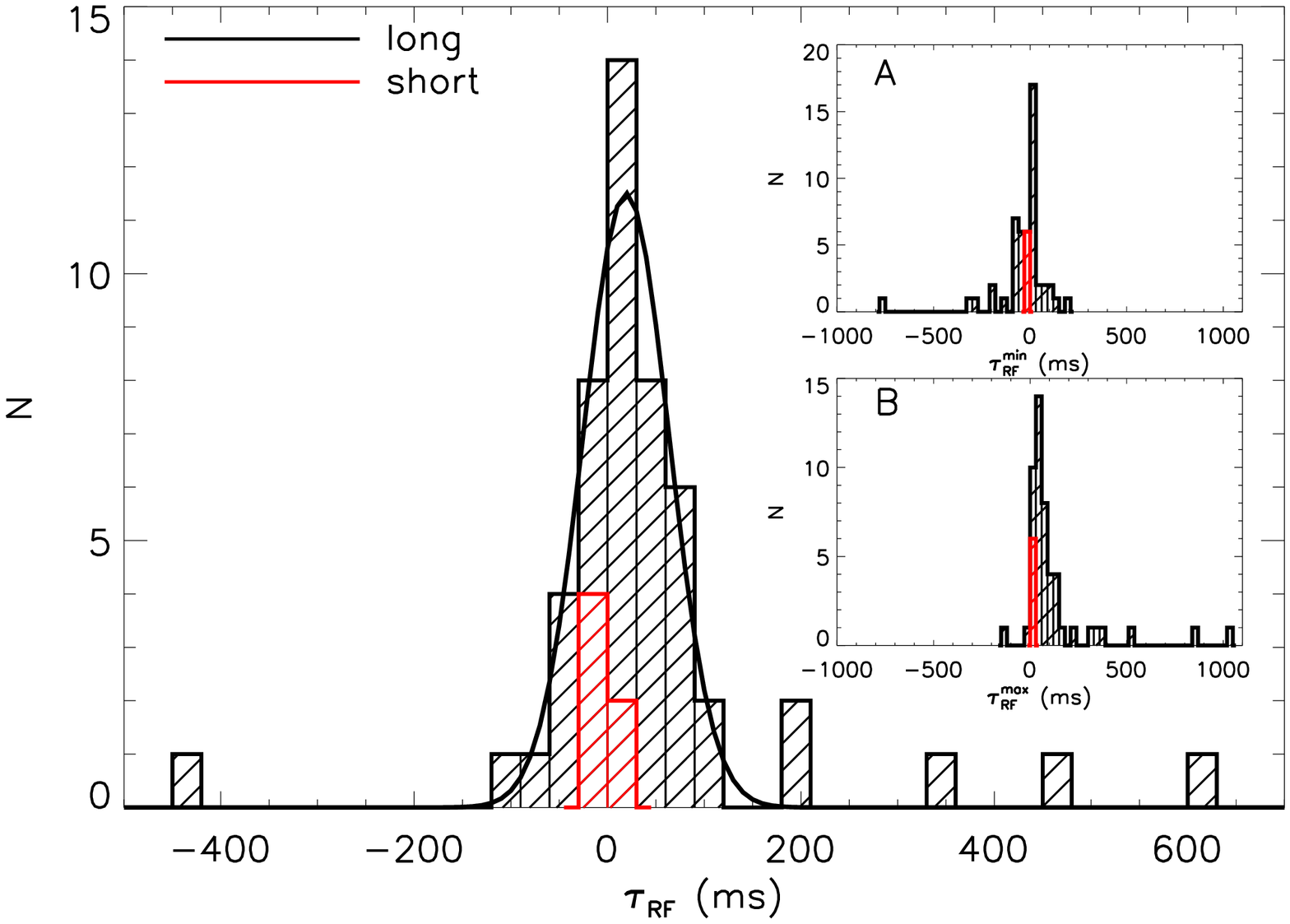}\\
\includegraphics[width= \hsize,clip]{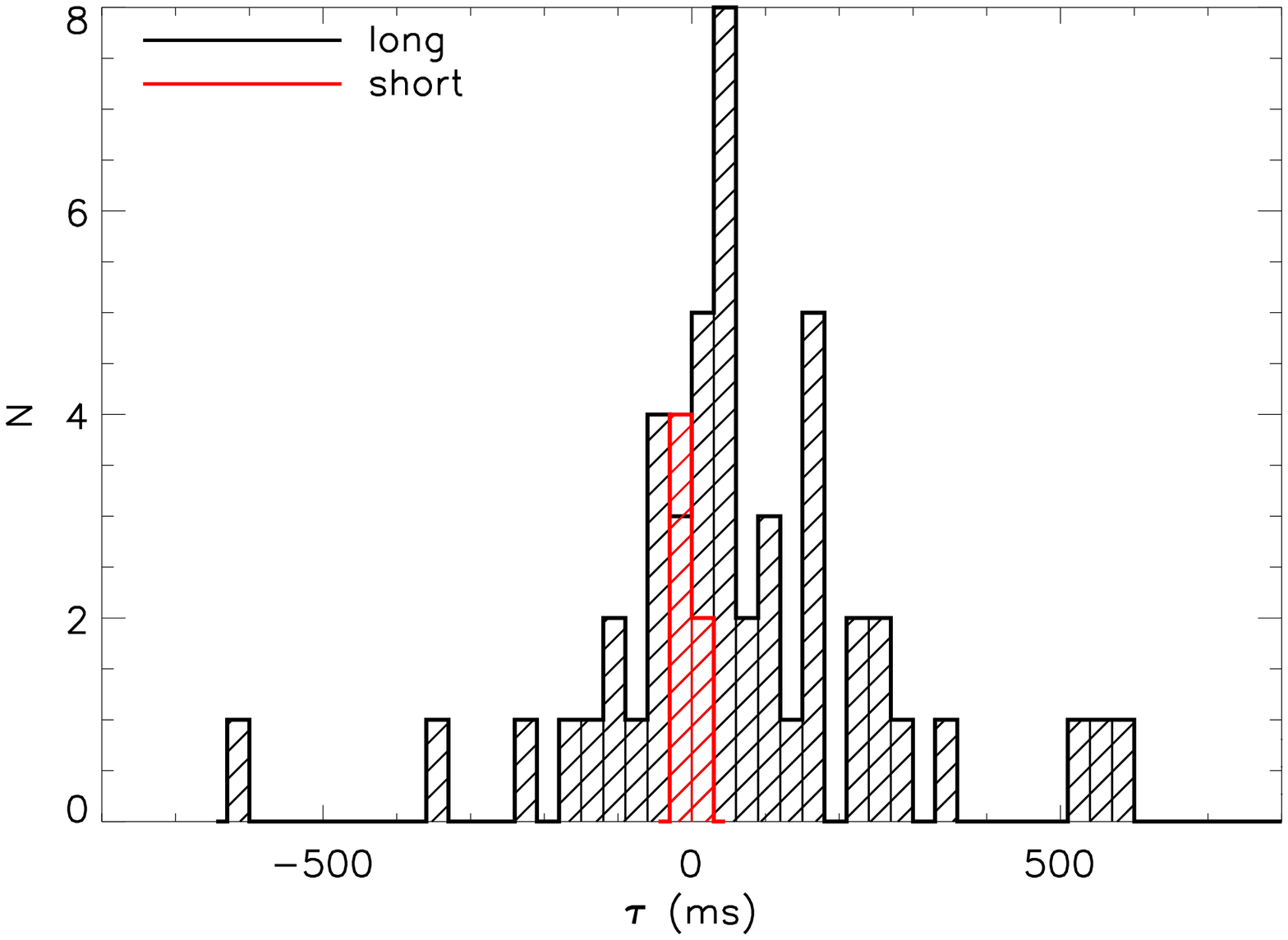}
\caption{Distribution of the spectral lags. Black: $50$ long GRBs from the BAT6 sample. Red: $6$ short GRBs from the S-BAT4 sample. Upper panel: rest frame spectral lags and their uncertainties for the GRBs of our samples. The lags are listed as in Table~\ref{table_tot}, from top to bottom. The black triangle corresponds to GRB 100816A. Middle panel: mean values of the rest frame spectral lags. The black solid line is a gaussian fit to the distribution for the long GRBs ($\mu=19.2$, $\sigma=44.4$, $N=11.5$). Inset A: minimum rest frame spectral lag, defined as $\tau_{\rm RF}^{\rm min}=\tau_{\rm RF}-\sigma_{\rm l,RF}$. Inset B: maximum rest frame spectral lag, defined as $\tau_{\rm RF}^{\rm max}=\tau_{\rm RF}+\sigma_{\rm r,RF}$. Lower panel: mean values of the observer-frame spectral lags.}
\label{fig_distr}
\end{figure}

For the short GRBs, we used the complete flux-limited sample of bright short GRBs with favorable observing conditions for ground-based optical follow-up aimed at redshift determination (hereafter S-BAT4) presented in \citet{2014arXiv1405.5131D}, comprising $11$ GRBs with redshift\footnote{We excluded GRB 080905A whose redshift has been questioned in \citet{2014arXiv1405.5131D}.}. 

The S-BAT4 sample is selected starting from all the \textit{Swift} GRBs classified as short by the BAT team refined analysis, namely all the GRBs with $T_{90} < 2$ s and those whose \textit{Swift}/BAT light curve shows a short-duration peak followed by a softer, long-lasting tail (the so-called ``extended emission'', with $T_{90} > 2$ s). The absence of spectral lag is a well-known feature for short GRBs \citep{2001grba.conf...40N,2006ApJ...643..266N}, and together with the prompt emission hardness ratio is often used as an indication for the nature of a burst with doubtful classification (when, e.g. $T_{90}\sim 2$ s). However, the calculation of spectral lag for short GRBs is not performed in a systematic way (the same energy bands, the same extraction method for the light curve) and is not available for all bursts. Specifically, only $\sim 50\%$ of the S-BAT4 sample has a reported spectral lag. The availability of only a fraction of spectral lags and the high dependence of its value on the reference energy bands indicate that the use of the S-BAT4 sample should not result in a strong bias towards short GRBs with negligible lags.

Since the $T_{90}$ distribution can vary among different instruments (e.g. due to the different energy bands), \citet{2013ApJ...764..179B} proposed that the dividing time between short and long GRBs should be reduced to $\sim 0.8$ s for \textit{Swift} bursts, with a possible contamination of long GRBs (i.e. it likely has a collapsar progenitor, \citealt{2006Natur.444.1010Z,2013ApJ...764..179B}) up to $50\%$ for $T_{90}\gtrsim 1$ s. \citet{2014arXiv1405.5131D} in their analysis identified $3$ possible long (i.e. collapsar) GRBs  out of $14$ with $T_{90}\gtrsim 1$ s, that is consistent with the expectations of \citet{2013ApJ...764..179B} within a factor $2$.

We computed the spectral lag for $7$\footnote{$4$ GRBs from the S-BAT4 sample - GRB 080123, GRB 090426, GRB 100117A and GRB 100625A - have been discarded from our analysis because there were not enough counts in one of the two energy bands - usually ch2 - to extract a significant value of the CCF for any choice of the time bin} out of $11$ short GRBs of the S-BAT4 sample: all the lags determined are consistent with zero within errors, $2$ with positive and $4$ with negative central values, with the exception of GRB 100816A that has a positive spectral lag (i.e. not consistent with zero within $1\,\sigma$). These results are reported in Table~\ref{table_tot}. However, GRB 100816A is one of the possible collapsar event \citep{2014arXiv1405.5131D}, thus we excluded it from our analysis and portrayed its results just for reference, ending up with $6$ short GRBs in our sample all consistent with zero.

\section{Results}\label{sect_r}

\subsection{Long GRBs}\label{ll}

Figure~\ref{fig_distr} shows the spectral lags and their uncertainties (upper panel, black points) and the distribution of the spectral lags of long GRBs (black hatched histogram) in the rest (middle panel) and in the observer frame (lower panel). The most extreme (negative and positive) spectral lags found and shown in fig.\ref{fig_distr} are for GRB 061021 and GRB 090926B, respectively. We also portrayed in the two insets of fig.~\ref{fig_distr} the distributions of the minimum rest frame spectral lag, defined as $\tau_{\rm RF}^{\rm min}=\tau_{\rm RF}-\sigma_{\rm l,RF}$ and of the maximum rest frame spectral lag, defined as $\tau_{\rm RF}^{\rm max}=\tau_{\rm RF}+\sigma_{\rm r,RF}$.


In order to account for the uncertainties on the spectral lag, that in some cases may be large (see fig.~\ref{fig_distr}, upper panel), we calculated the moments of the distribution from $10,000$ distributions of spectral lags obtained from the original one by assuming that each spectral lag is normally distributed around the calculated value, with a standard deviation equal to its uncertainty. We found that the mean (median) value of the distribution for long GRBs is $\langle \tau_{RF}^L \rangle=(43.0\pm17.8)$ ms ($\bar{\tau}_{RF}^L=(24.9\pm7.1)$ ms), and the standard deviation is $\sigma^L=(186.3\pm42.1)$ ms (hereafter, the superscript L (S) stands for long (short) GRBs).


\begin{figure*}
\centering
\includegraphics[width= 0.45 \hsize,clip]{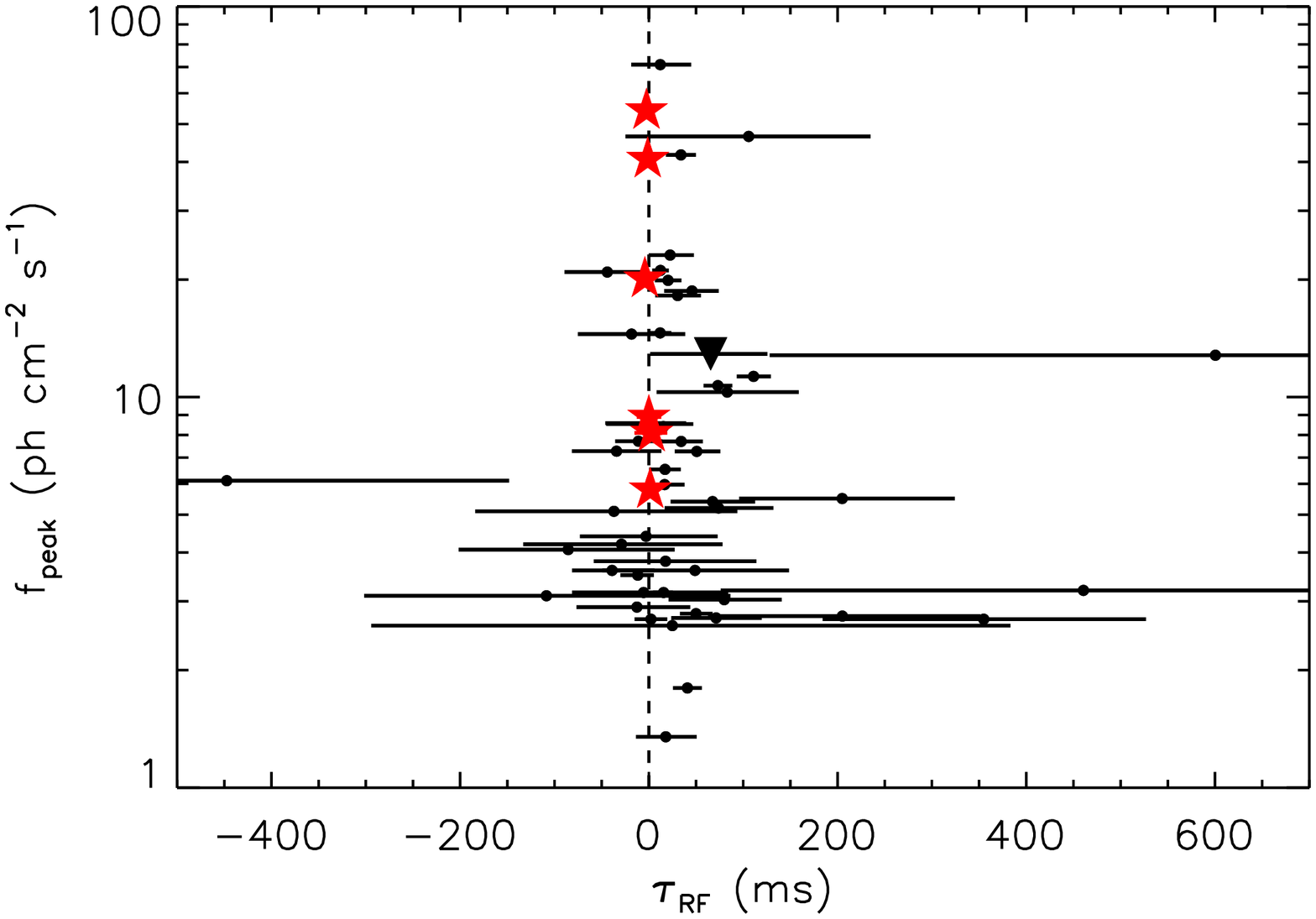}
\includegraphics[width= 0.45 \hsize,clip]{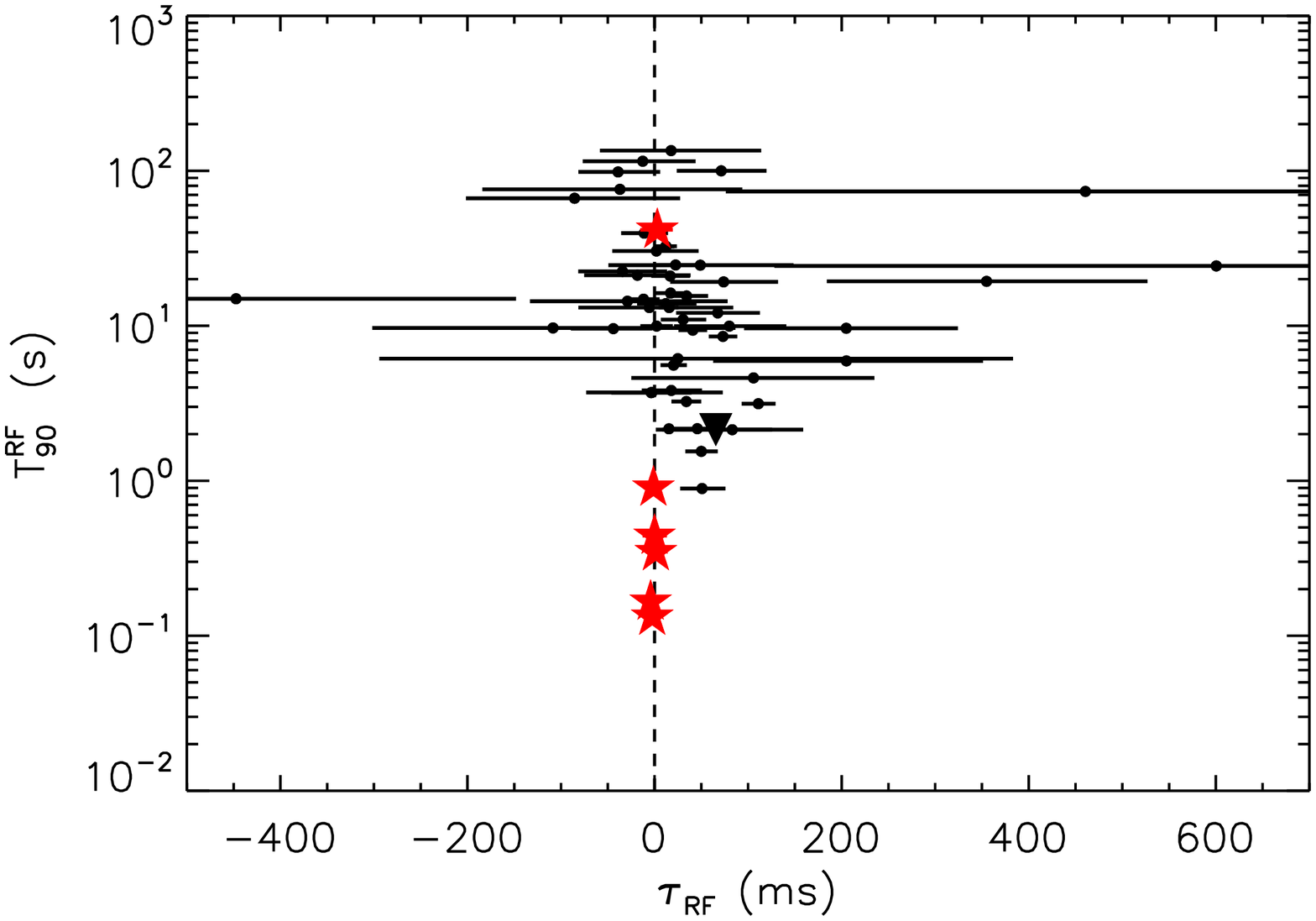}
\caption{Photon peak flux in the $15-150$ keV energy band (left panel) and rest-frame $T_{90}$ in the same energy band (right panel) versus rest-frame spectral lag. Black points: $50$ long GRBs from the BAT6 sample. Red stars: $6$ short GRBs from the S-BAT4 sample. The black triangle corresponds to GRB 100816A.}
\label{fig_pT}
\end{figure*}

Since according to our definition the spectral lag is a time-integrated property, it is possible that when the CCF versus time delay shows a broader peak also the uncertainty associated to the lag is larger. This could be due to the presence of different spectral lags during different peaks in a single burst \citep{2008ApJ...677L..81H}. We investigated this possibility by simulating light curves composed by several peaks each with slightly different values of the lag (i.e. the difference in the lag values is comparable to light curve bin size). We found that the overall lag estimated through the CCF over the entire light curve is closer to the value of the lag of the leading peak in the simulated light curve, and the leading peak is resolved with less precision than for single lag light curves. The detailed study of this possibility will be presented in a forthcoming paper (Lazzati et al., in prep.).

\subsection{Short GRBs}

Figure~\ref{fig_distr} shows the spectral lags and their uncertainties (upper panel, red stars) and the distribution of the spectral lags of short GRBs (red hatched histogram) in the rest (middle panel) and in the observer frame (lower panel) compared to the long GRB one. Similarly, the two insets of fig.~\ref{fig_distr} compare their distributions of the minimum rest frame spectral lag and of the maximum rest frame spectral lag. The moments of the distribution for short GRBs are derived from simulations that account for the uncertainties, as already described for long events. Short GRBs have spectral lag values centered around zero, with a mean (median) value of the distribution $\langle \tau_{RF}^S \rangle=(-0.61\pm3.87)$ ms ($\bar{\tau}_{RF}^S=(0.01\pm2.77)$ ms), and standard deviation $\sigma^S=(8.98\pm3.94)$ ms that is much narrower than for long GRBs. 

When we compare the two distributions for long and short GRBs, we find no statistical evidence that they are drawn from two different populations. Specifically, if we perform a Kolmogorov-Smirnov (KS) test for each of the $10,000$ distributions of spectral lags obtained accounting for the uncertainties, the mean probability that the two samples are drawn from the same population is $4.1\%$. Albeit the characterization of the spectral lag of short GRBs might still suffer from the small number of events, we cannot exclude at more than $2\,\sigma$ that the two populations have similar lags distributions.

If we compare the distributions of the spectral lag in the observer frame for long and short GRBs (see fig.~\ref{fig_distr}, lower panel), we find that the mean values are more separated ($\langle \tau^L \rangle=(102.2\pm38.1)$ ms and $\langle \tau^S \rangle=(-0.73\pm7.14)$ ms), and with broader distributions ($\sigma^L=(375.1\pm69.6)$ ms and $\sigma^S=(16.5\pm7.5)$ ms) as a consequence of the different redshift distributions of the short and long GRBs: the average redshift for the long GRB sample is $\langle z^L \rangle = 1.84$ whereas for the short GRB sample is $\langle z^S \rangle = 0.85$ \citep{2012ApJ...749...68S,2014arXiv1405.5131D}. The KS test gives a probability of $2\%$ that they are drawn from the same population. This demonstrates the importance of performing the lag analysis in the rest frame when the aim is to compare short and long GRBs and when the lag is associated with a rest frame property such as the luminosity.  

We also compared the two distributions normalising the spectral lag to the $T_{90}$. Long and short GRBs are more similar when accounting for their different timescales (see e.g. \citealt{2010ApJ...725..225G}): their lags cluster around a similar value ($\langle (\tau/T_{90})^L\rangle=(1.92\pm1.85)$ ms and $\langle (\tau/T_{90})^S\rangle=(-4.04\pm8.62)$ ms) with similar spread ($\sigma^L=(21.3\pm7.6)$ ms and $\sigma^S=(19.5\pm9.7)$ ms). 

We investigated the possible dependence of the lag values and their uncertainty on other burst properties as observed by BAT (both for long and short GRBs), such as the photon peak flux in the $15-150$ keV energy band or the rest-frame $T_{90}$ (both computed in the same observed $15-150$ keV energy band)\footnote{For GRBs until Dec 2009 we referred to the 2nd \textit{Swift}/BAT catalogue \citep{2011ApJS..195....2S}, while for events occuring later than this date, to the refined analysis GCN circulars of the \textit{Swift}/BAT team (http:gcn.gsfc.nasa.gov/gcn3.archive.html).}.  Short and long GRBs have similar peak fluxes in $15-150$ keV energy band, as a result of the selection criteria of the two samples. Figure~\ref{fig_pT} shows that there is no correlation of these quantities with $\tau_{RF}$: spectral lags consistent with zero are present in bursts with either high (low) peak flux and long (short) duration. The only short burst with an extended emission in our sample, GRB 070714B, has a negligible lag, as the other short bursts with duration $\lesssim 2$ s. Since the dividing line between short and long GRBs on the basis of the $T_{90}$ is established in the observer frame, we searched for any evolution of the spectral lag on the $T_{90}$ in the observer frame. The situation is similar to the one in the rest frame: the shortest long bursts do not have a smaller lag compared to the longest ones.

\subsection{Lag estimates with Fermi data}

In order to verify if our results on the spectral lags and their uncertainty are somehow related to the BAT instrument, we performed the same analysis on the GRBs in our sample that were also detected by the Gamma Burst Monitor (GBM) on board the \textit{Fermi} satellite ($11$ long GRBs and $1$ short GRB, see Table~\ref{table_fermi}). We used the standard GBM analysis tools (RMFIT v. 4.3.2) to extract the light curves from the time tagged events (TTE) files in the same rest frame energy bands adopted for the analysis of the BAT data. We also fitted the background with a polynomial function by interpolating two background intervals before and after the bursts. In all cases we used only the NaI data ($15$ keV $-1$ MeV), which have enough energy coverage to encompass the rest frame ch1 and ch2 energy bands discussed in Section~\ref{sect_m}. 

\begin{figure}
\centering
\includegraphics[width= \hsize,clip]{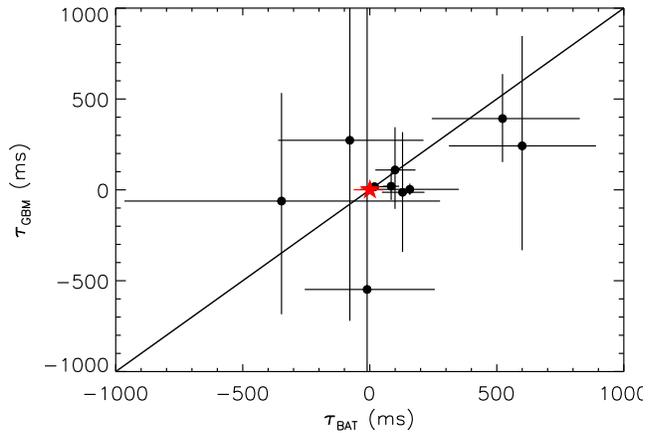}
\caption{Observer-frame spectral lag measured with \textit{Swift}/BAT and \textit{Fermi}/GBM data for the $11$ long GRBs (black points) and the $1$ short GRB (red star) observed by both instruments. The black line corresponds to $\tau_{\rm BAT}=\tau_{\rm GBM}$.}
\label{fig_fermi}
\end{figure}

Figure~\ref{fig_fermi} shows the comparison of the rest-frame spectral lag computed with the \textit{Fermi} and \textit{Swift} data. The two estimates are consistent within their errors, and on average the lag computed with the \textit{Fermi} data leads to a slightly larger uncertainty due to the different background subtraction method of the BAT and GBM instruments.

\subsection{Lag-luminosity correlation}

\begin{figure*}
\centering
\includegraphics[width= \hsize,clip]{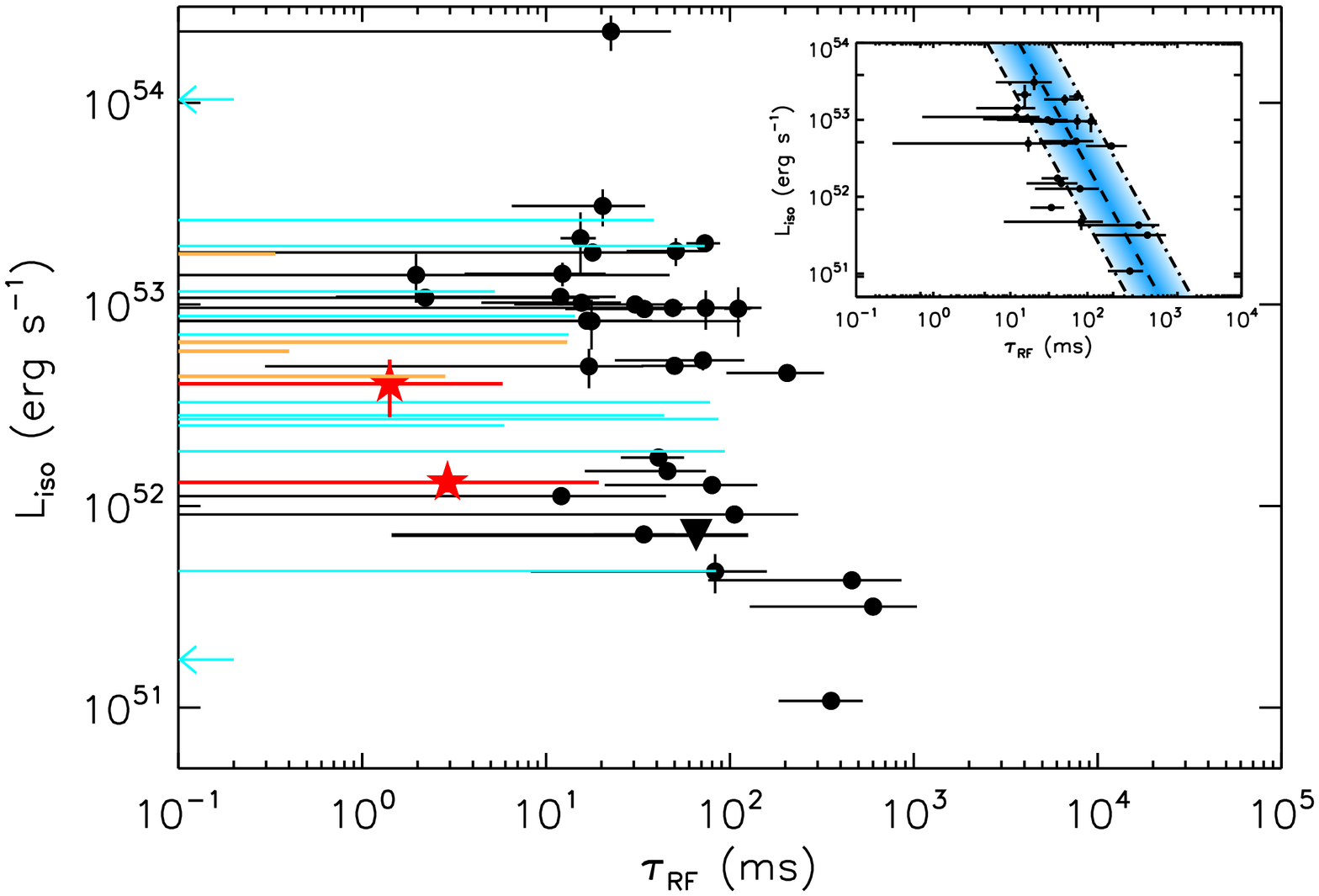}
\caption{Peak luminosity $L_{\rm iso}$ as a function of the rest-frame spectral lag. Black points: long GRBs with positive central value of the spectral lag ($23$ with positive lag and $9$ with positive lag consistent with zero within errors). Cyan points: long GRBs with negative central value of the spectral lag ($2$ with negative lag, marked as left arrows, and $11$ with negative lag consistent with zero within errors). Red stars: short GRBs with positive central value of the spectral lag ($2$ with positive lag consistent with zero within errors). Orange stars: short GRBs with negative central value of the spectral lag ($4$ with negative lag consistent with zero within errors). The black triangle corresponds to GRB 100816A. Inset: lag-luminosity anti-correlation for the $23$ long GRBs with positive lag. The black dashed line is the best fit to the data: ${\rm log}[L_{\rm iso}/({\rm 10^{52} erg\,s^{-1}})]=(0.42\pm0.11)+(-1.79\pm0.03)\,{\rm log}[\tau_{RF}/{\rm 100\,ms}]$, and the blue area marks the $1-\sigma$ region around the best fit.}
\label{fig_lag_lum}
\end{figure*}

We considered all the GRBs with measured lags in our samples (long and short) that also have an estimate of the bolometric isotropic luminosity $L_{\rm iso}$ to investigate the relation between the spectral lag and the GRB luminosity, namely $45$ long GRBs and $6$ short GRBs. For the values of $L_{\rm iso}$ and its definition we refer to \citet{2012MNRAS.421.1256N}.

In analogy with previous works \citep{2010ApJ...711.1073U,2012MNRAS.419..614U}, we first restricted our analysis to all long GRBs with positive spectral lag ($23$; $51\%$ of the sample): we found that the luminosity significantly anti-correlates with the spectral lag (Pearson correlation coefficient $r=-0.68$, null-hypothesis probability $P=3.8\times10^{-4}$). The best linear fit to the ${\rm log}(L_{\rm iso})-{\rm log}[\tau_{RF}]$ correlation that accounts for the statistical uncertainties on both axes yields: ${\rm log}[L_{\rm iso}/({\rm 10^{52} erg\,s^{-1}})]=(0.42\pm0.11)+(-1.79\pm0.03)\,{\rm log}[\tau_{RF}/{\rm 100\,ms}]$ (see fig.~\ref{fig_lag_lum}, inset). The scatter perpendicular to the correlation is modelled with a Gaussian with standard deviation $\sigma=0.65$.

However, the restriction to the GRBs with a spectral lag significantly ($1\,\sigma$) greater than zero and the consequent exclusion of about a half of the total sample introduces a bias: since the physical origin of the spectral lag and of this correlation is not well understood, there is no a priori reason to consider lags consistent with zero or negative as spurious. We therefore added to the lag-luminosity plane also the $9$ long GRBs with positive lag but consistent with zero within errors and  the $11$ long GRBs with negative lag but consistent with zero within errors. The results are portrayed in fig.~\ref{fig_lag_lum}. No correlation between $L_{\rm iso}$ and the spectral lag is anymore apparent. There are also two long GRBs with negative lag (GRB 061021 and GRB 080721; cyan arrows in fig.~\ref{fig_lag_lum}, see fig.~\ref{fig_061021} where the lag computation for GRB 061021 is explicitly shown): they correspond to an high (GRB 080721) and a low (GRB 061021) luminosity event. 

\begin{figure}
\centering
\includegraphics[width= \hsize,clip]{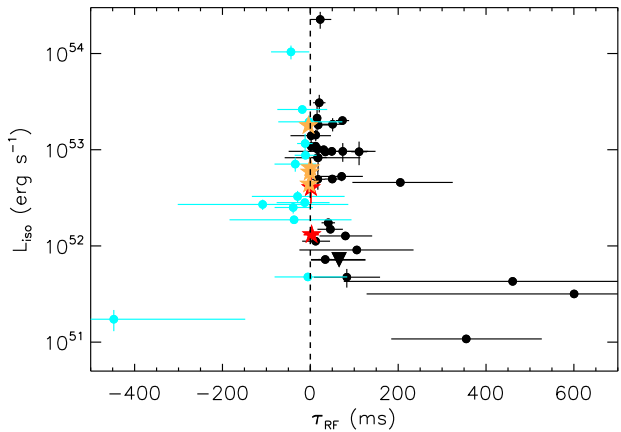}
\caption{Peak luminosity $L_{\rm iso}$ as a function of the rest-frame spectral lag. Black points: long GRBs with positive central value of the spectral lag ($23$ with positive lag and $9$ with positive lag consistent with zero within errors). Cyan points: long GRBs with negative central value of the spectral lag ($2$ with negative lag and $11$ with negative lag consistent with zero within errors). Red stars: short GRBs with positive central value of the spectral lag ($2$ with positive lag consistent with zero within errors). Orange stars: short GRBs with negative central value of the spectral lag ($4$ with negative lag consistent with zero within errors). The black triangle corresponds to GRB 100816A.}
\label{fig_lag_lum_lin}
\end{figure}

Figure~\ref{fig_lag_lum_lin} shows the luminosity versus the spectral lag on a linear scale. GRBs with high luminosity seem to have smaller lags in absolute value, while low luminosity GRBs span a wider range of spectral lags. However, we divided the long GRBs of our sample in two groups with luminosity above and below $10^{52}$ erg s$^{-1}$ and we compared their spectral lag distributions accounting for the uncertainties as described in Section~\ref{ll}.  A KS test results in a probability $P=42\%$ that they are drawn from the same population. Similarly, the two populations of long GRBs with positive spectral lag and the ones with lag consistent with zero are not statistically different ($P=12\%$). We also compared the luminosity distributions for the two groups (positive and negligible spectral lags) and we found a probability $P=65\%$ that they are drawn from the same population.

We investigated the lag-luminosity relation also for the short GRBs of our sample\footnote{For the values of $L_{\rm iso}$ and its definition we refer to \citet{2014arXiv1405.5131D}.} ($6$; see fig.~\ref{fig_lag_lum}): they do not occupy a separate region of the lag-luminosity plane when compared to the total sample of long GRBs, because neither their distribution of the spectral lag is significantly different from the long GRB one, as discussed in the previous section, nor the luminosity distributions (KS probability $P=38\%$). The small number of short GRBs does not allow us to draw any firm conclusion about the presence of a boundary for short GRBs (see fig.~\ref{fig_lag_lum_lin}). GRB 100816A is consistent with the spectral lags of long GRBs with comparable luminosity.

\section{Conclusions}\label{sect_c}

We extracted the spectral lag in the cosmological rest frame between two fixed rest frame energy bands from the BAT data of the \textit{Swift} satellite. We considered two samples of $50$ long  and $6$ short GRBs from the BAT6 \citep{2012ApJ...749...68S} and S-BAT4 \citep{2014arXiv1405.5131D} complete samples, respectively. With the background subtracted light curves we computed the discrete CCF and fitted it with an asymmetric Gaussian model to search for its global maximum that, by definition, corresponds to the spectral lag $\tau_{RF}$. We accounted for the errors on the data points through a Monte Carlo method to estimate the uncertainty on $\tau_{RF}$. 

We found that:
\begin{itemize}
\item the spectral lag between the chosen rest frame energy bands for long GRBs  is significantly (within $1\,\sigma$) greater than zero in most cases ($50\%$). However an equally large fraction ($50\%$) of them are consistent with zero or negative within errors;
\item short GRBs have in all cases limited or no lag in the same rest frame energy bands. GRB 100816A has a significantly positive lag, however it is likely a short duration GRB with a collapsar progenitor \citep{2014arXiv1405.5131D}; 
\item the distribution of the spectral lags for short GRBs is peaked at a smaller value than the long GRB distribution. However, there is no stronger than $2\,\sigma$ statistical indication that the spectral lags of short and long GRBs are drawn from two different populations;
\item the estimate of the time-integrated lag is limited by the signal to noise ratio of the light curves, and it is not determined by the duration or peak flux of the event: GRBs (either long or short) with large or small peak flux can have a positive or null lag;
\item the lag estimates we derived from the BAT data are consistent with those derived by similar analysis of the \textit{Fermi}/GBM data;
\item for those GRBs of our samples with peak luminosity $L_{\rm iso}$ ($45$ long and $6$ short GRBs) we investigated the lag-luminosity correlation. We recover the correlation when considering only long GRBs with positive lag, confirming previous results \citep{2010ApJ...711.1073U,2012MNRAS.419..614U}. However, when we include in the lag-luminosity plane also the long GRBs with lag consistent with zero, the correlation is weakened and it appears that the left-hand side of the $L_{\rm iso}-\tau_{RF}$ correlation is filled with bursts. 
\end{itemize}


The main conclusions that we draw from our analysis is that the time-integrated spectral lag as a tool to distinguish between short and long GRBs might not be as definite as thought before: the existence of a large fraction of long GRBs with a lag consistent with zero makes it challenging to classify all those ambiguous GRBs (e.g. long GRBs with a duration shorter than $2$ s in the observer frame because they are at high redshift) as short only because they have a null lag.

The estimate of a null lag or of a lag consistent with zero is not connected to the observed properties (rest frame duration and peak flux) of the bursts tested in this work. Though the overall time-integrated spectral lag and its uncertainty can still be dependent upon the light curve structure, spectral lags consistent with zero are reliable estimates as well as (significative) positive lags and cannot be excluded in the estimate of the lag-luminosity correlation. Indeed when including all the bursts with measured lags, the lag-luminosity plane fills on the left hand side of the previously known lag-luminosity correlation that now appears as a boundary in this plane, challenging also the possible use of this relation for cosmological purposes. The possibility that this boundary is still affected by biases is beyond the scope of the present work. Similar conclusions have been found by \citet{2013A&A...557A.100H} about the $E_{\rm pk}-E_{\rm iso}$ correlation.

Long and short GRBs occupy only slightly different positions in this plane (albeit the characterization of the spectral lag of short GRBs might still suffer from the small number of events with measured redshift and well determined prompt emission spectrum - which guarantees the estimate of $L_{\rm iso}$). Therefore, the lag-luminosity correlation is questioned by our findings.

\section*{Acknowledgements}
The authors thank the anonymous referee for his/her useful comments, and D. Burrows for valuable suggestions. The authors acknowledge support from ASI-INAF I/088/06/0 and PRIN-INAF 1.05.01.09.15 grants. DB acknowledge the support of the Australian Research Council through grant DP110102034.

\begin{table*}
\caption{Spectral lags for the $50$ long (upper part), the $6$ short (lower part) GRBs of our samples and GRB 100816A. GRB name, redshift (z), temporal resolution (bin), left ($t_l$) and right ($t_r$) boundaries of the time interval over which the spectral lag is computed, spectral lag in the observer frame ($\tau$), left ($\sigma_l$) and right ($\sigma_r$) uncertainties.}
\label{table_tot}
\begin{tabular}{cccccccc}
\hline
\hline
GRB name & z & bin (ms) & $t_l$ (s) & $t_r$ (s) & $\tau$ (ms) & $\sigma_l$ (ms) & $\sigma_r$ (ms)\\
\hline  																	        		      
050318 & $     1.44$ & $  64$ & $     23.0$ & $     50.0$ & $   -13.66$ & $   184.88$ & $   218.76$ \\
050401 & $     2.90$ & $  64$ & $     23.0$ & $     29.0$ & $   285.19$ & $    59.05$ & $    59.14$ \\
050525A & $     0.61$ & $  16$ & $     -1.0$ & $      9.0$ & $    54.72$ & $    25.42$ & $    25.59$ \\
050802 & $     1.71$ & $ 256$ & $     -5.0$ & $     20.0$ & $   555.80$ & $   386.11$ & $   395.90$ \\
050922C & $     2.20$ & $  16$ & $     -3.0$ & $      3.0$ & $   162.52$ & $    74.74$ & $    79.50$ \\
060206 & $     4.05$ & $  16$ & $     -1.5$ & $      8.0$ & $   252.40$ & $    85.65$ & $    88.18$ \\
060210 & $     3.91$ & $ 128$ & $     -3.3$ & $      5.0$ & $   349.99$ & $   233.64$ & $   237.12$ \\
060306 & $     1.55$ & $  32$ & $      0.0$ & $      5.0$ & $    42.56$ & $    51.17$ & $    53.73$ \\
060814 & $     1.92$ & $  64$ & $     10.0$ & $     25.0$ & $  -100.01$ & $   138.04$ & $   138.73$ \\
060908 & $     1.88$ & $  32$ & $    -11.0$ & $      4.0$ & $   230.04$ & $   169.95$ & $   175.42$ \\
060912A & $     0.94$ & $  64$ & $     -1.0$ & $      5.0$ & $    -7.09$ & $    82.58$ & $    83.49$ \\
060927 & $     5.47$ & $  32$ & $     -2.0$ & $      8.0$ & $    14.26$ & $   111.90$ & $   111.69$ \\
061007 & $     1.26$ & $   4$ & $     24.0$ & $     65.0$ & $    27.05$ & $    25.42$ & $    26.88$ \\
061021 & $     0.35$ & $ 512$ & $     -0.5$ & $     15.0$ & $  -603.94$ & $   416.22$ & $   403.94$ \\
061121 & $     1.31$ & $   4$ & $     60.5$ & $     80.5$ & $    28.36$ & $    20.02$ & $    20.25$ \\
061222A & $     2.09$ & $  64$ & $     25.0$ & $     30.0$ & $     6.07$ & $   145.67$ & $   139.01$ \\
070306 & $     1.50$ & $  32$ & $     90.0$ & $    118.0$ & $  -213.78$ & $   290.08$ & $   281.92$ \\
070521 & $     1.35$ & $  16$ & $     15.0$ & $     40.0$ & $    40.20$ & $    39.51$ & $    39.07$ \\
071020 & $     2.15$ & $   4$ & $     -3.0$ & $      1.0$ & $    48.47$ & $    10.70$ & $    10.24$ \\
071117 & $     1.33$ & $  16$ & $     -1.0$ & $      3.0$ & $   258.54$ & $    41.21$ & $    42.58$ \\
080319B & $     0.94$ & $   4$ & $     -3.0$ & $     58.0$ & $    30.29$ & $    21.67$ & $    19.18$ \\
080319C & $     1.95$ & $  32$ & $     -1.0$ & $     13.5$ & $   217.82$ & $   168.48$ & $   171.20$ \\
080413B & $     1.10$ & $  32$ & $     -1.5$ & $      5.0$ & $    96.00$ & $    61.91$ & $    59.56$ \\
080430 & $     0.77$ & $ 256$ & $     -1.5$ & $     13.0$ & $    44.04$ & $   564.87$ & $   634.35$ \\
080603B & $     2.69$ & $  16$ & $     -0.5$ & $      5.0$ & $   -43.59$ & $    67.38$ & $    63.01$ \\
080605 & $     1.64$ & $   8$ & $     -5.5$ & $     16.0$ & $    53.65$ & $    36.46$ & $    37.38$ \\
080607 & $     3.04$ & $   8$ & $     -6.0$ & $     12.0$ & $    90.99$ & $    91.44$ & $   101.78$ \\
080721 & $     2.59$ & $  64$ & $     -3.5$ & $      8.5$ & $  -158.16$ & $   162.73$ & $   149.69$ \\
080804 & $     2.20$ & $ 256$ & $     -5.0$ & $     20.0$ & $  -347.40$ & $   618.25$ & $   623.99$ \\
080916A & $     0.69$ & $ 128$ & $     -5.0$ & $     10.0$ & $   599.82$ & $   288.57$ & $   290.73$ \\
081121 & $     2.51$ & $ 256$ & $      0.0$ & $     20.0$ & $   -10.41$ & $   245.62$ & $   266.41$ \\
081203A & $     2.10$ & $ 128$ & $     25.0$ & $     40.0$ & $   -39.23$ & $   198.37$ & $   175.09$ \\
081221 & $     2.26$ & $  16$ & $     15.0$ & $     40.0$ & $    99.44$ & $    77.55$ & $    80.56$ \\
081222 & $     2.77$ & $  64$ & $     -1.0$ & $     16.0$ & $   129.02$ & $    81.04$ & $    86.36$ \\
090102 & $     1.55$ & $ 256$ & $    -15.0$ & $     20.0$ & $   522.53$ & $   278.44$ & $   304.17$ \\
090201 & $     2.10$ & $  64$ & $      0.0$ & $     50.0$ & $   -56.92$ & $   175.92$ & $   176.01$ \\
090424 & $     0.54$ & $  16$ & $     -1.0$ & $      5.0$ & $    18.62$ & $    47.22$ & $    50.44$ \\
090709A & $     1.80$ & $  64$ & $    -10.0$ & $     70.0$ & $   -31.00$ & $    68.71$ & $    71.05$ \\
090715B & $     3.00$ & $  16$ & $     -5.0$ & $     21.0$ & $    70.66$ & $   304.24$ & $   385.39$ \\
090812 & $     2.45$ & $ 256$ & $     -7.0$ & $     41.0$ & $   168.71$ & $   338.84$ & $   343.29$ \\
090926B & $     1.24$ & $ 256$ & $    -22.0$ & $     36.0$ & $  1031.73$ & $   861.13$ & $   887.57$ \\
091018 & $     0.97$ & $  64$ & $     -0.3$ & $      3.0$ & $   163.65$ & $   147.37$ & $   149.05$ \\
091020 & $     1.71$ & $ 128$ & $     -2.5$ & $     14.0$ & $   -78.58$ & $   282.06$ & $   290.03$ \\
091127 & $     0.49$ & $  64$ & $     -1.0$ & $      2.0$ & $   157.64$ & $   194.65$ & $   192.49$ \\
091208B & $     1.06$ & $  64$ & $      7.5$ & $     10.5$ & $    84.20$ & $    31.61$ & $    31.60$ \\
100615A & $     1.40$ & $ 128$ & $     -5.0$ & $     40.0$ & $   162.03$ & $   106.60$ & $   108.27$ \\
100621A & $     0.54$ & $ 256$ & $     -6.5$ & $     40.5$ & $   924.74$ & $   727.39$ & $   677.68$ \\
100728B & $     2.11$ & $ 256$ & $     -5.0$ & $      7.0$ & $  -115.00$ & $   456.44$ & $   406.26$ \\
110205A & $     2.22$ & $  64$ & $    118.0$ & $    294.0$ & $  -125.63$ & $   136.21$ & $   144.66$ \\
110503A & $     1.61$ & $  32$ & $     -2.0$ & $      8.0$ & $    46.77$ & $    82.15$ & $    85.65$ \\
\hline
\hline
051221A & $     0.55$ & $   4$ & $     -0.3$ & $      0.5$ & $    -1.85$ & $     2.32$ & $     2.47$ \\
070714B & $     0.92$ & $  16$ & $     -1.0$ & $      2.0$ & $     5.58$ & $    35.01$ & $    31.56$ \\
090510 & $     0.90$ & $   8$ & $     -0.2$ & $      0.5$ & $    -7.99$ & $     8.40$ & $     8.63$ \\
101219A & $     0.72$ & $  16$ & $     -0.5$ & $      1.0$ & $    -0.02$ & $    21.77$ & $    22.42$ \\
111117A & $     1.30$ & $  16$ & $     -0.5$ & $      1.0$ & $     3.24$ & $    10.70$ & $    10.10$ \\
130603B & $     0.36$ & $   8$ & $     -0.3$ & $      0.3$ & $    -3.44$ & $     5.58$ & $     7.27$ \\
\hline
100816A & $     0.81$ & $  64$ & $    -2.00$ & $     3.00$ & $   118.17$ & $   115.56$ & $   108.60$ \\
\end{tabular}	
\end{table*}	 	 	 

\begin{table}
\caption{Spectral lags for the GRBs of our samples obtained using \textit{Fermi}/GBM data. GRB name, redshift (z), temporal resolution (bin), spectral lag in the observer frame $\tau$, left ($\sigma_l$) and right ($\sigma_r$) uncertainties.}
\label{table_fermi}
\begin{tabular}{cccccc}
\hline
\hline
GRB name & z & bin (ms) & $\tau$ (ms) & $\sigma_l$ (ms) & $\sigma_r$ (ms)\\
\hline  																	        		      
080804 & $     2.20$ & $ 256$ & $   -61.63$ & $   622.68$ & $   594.77$ \\
080916A & $     0.69$ & $ 128$ & $   242.04$ & $   574.08$ & $   604.30$ \\
081121 & $     2.51$ & $ 528$ & $  -548.14$ & $  2667.28$ & $  2136.56$ \\
081221 & $     2.26$ & $  32$ & $   109.42$ & $   213.98$ & $   235.29$ \\
081222 & $     2.77$ & $  64$ & $   -13.41$ & $   327.90$ & $   331.01$ \\
090102 & $     1.55$ & $ 128$ & $   392.29$ & $   239.42$ & $   245.04$ \\
090424 & $     0.54$ & $  16$ & $    17.35$ & $     9.78$ & $    10.33$ \\
090510 & $     0.90$ & $  32$ & $     0.56$ & $    64.56$ & $    53.34$ \\
090926B & $     1.24$ & $ 512$ & $  1820.21$ & $  2806.47$ & $  2828.40$ \\
091020 & $     1.71$ & $ 256$ & $   272.76$ & $   993.40$ & $  1000.52$ \\
091127 & $     0.49$ & $  64$ & $     2.87$ & $    30.43$ & $    31.13$ \\
091208B & $     1.06$ & $  64$ & $    19.84$ & $    75.36$ & $    76.31$ \\
\hline	
\end{tabular}	
\end{table}	 	 	 

\label{lastpage}

\end{document}